\documentclass[twoside,twocolumn,english,aps,prl,epsf,amssymb]{revtex4}
\usepackage[T1]{fontenc}
\usepackage[latin1]{inputenc}
\usepackage{babel}
\usepackage{graphics}
\usepackage{epsfig}
\makeatletter

\providecommand{\LyX}{L\kern-.1667em\lower.25em\hbox{Y}\kern-.125emX\@}

\makeatother
\begin{document}

\draft
\input{epsf.tex}

\newcommand{\tT}{\mbox{\scriptsize\sf T}} 

\title{Improved message passing for inference in densely connected systems }

\author{Juan P.~Neirotti and David~Saad}

\address{The Neural Computing Research Group, Aston University, Birmingham
B4 7ET, UK.}

\begin{abstract}
An improved inference method for densely connected systems is
presented. The approach is based on passing condensed messages between
variables, representing macroscopic averages of microscopic
messages. We extend previous work that showed promising results in
cases where the solution space is contiguous to cases where
fragmentation occurs. We apply the method to the signal detection
problem of Code Division Multiple Access (CDMA) for demonstrating its
potential. A highly efficient practical algorithm is also derived on
the basis of insight gained from the analysis.
\end{abstract}
\maketitle

Graphical models (Bayes belief networks) provide a powerful framework
for modelling statistical dependencies between
variables~\cite{Pearl,Jensen,MacKay_book}. They play an essential role
in devising a principled probabilistic framework for inference in a
broad range of applications from medical expert systems, to decoders
in telecommunication systems.

Message passing techniques are typically used for inference in
graphical models that can be represented by a sparse graph with a few
(typically long) loops. They are aimed at obtaining (pseudo) posterior
estimates for the system's variables by iteratively passing messages
(locally calculated conditional probabilities) between
variables. Iterative message passing of this type is guaranteed to
converge to the globally correct estimate when the system is
tree-like; there are no such guarantees for systems with loops even in
the case of large loops and a local tree-like structure (although
message passing techniques have been used successfully in loopy
systems, supported by some limited theory~\cite{weiss}). A clear link
has been established between certain message passing algorithms and
well known methods of statistical mechanics~\cite{MFA_book} such as
the Bethe approximation~\cite{TAP_EPL,YFW}.

These inherent limitations seem to prevent the use of message passing
techniques in densely connected systems due to their high
connectivity, implying an exponentially growing cost, and an
exponential number of loops. However, an exciting new approach has
been recently suggested~\cite{Kabashima_CDMA} for extending Belief
Propagation (BP) techniques~\cite{Pearl,Jensen,MacKay_book} to densely
connected systems. In this approach, messages are grouped together,
giving rise to a macroscopic random variable, drawn from a Gaussian
distribution of varying mean and variance for each of the nodes. The
technique has been successfully applied to signal detection in Code
Division Multiple Access (CDMA) problems and the results reported are
competitive with those of other state of the art techniques. However,
the current approach has some inherent
limitations~\cite{Kabashima_CDMA}, presumably due to its similarity to
the replica symmetric solution in equivalent Ising spin
models~\cite{MPV,Nishimori_book}.

In a separate recent development~\cite{MPZ}, the
replica-symmetric-equivalent BP has been extended to Survey
Propagation (SP), which corresponds to one-step replica symmetry
breaking in diluted systems. This new algorithm, motivated by the
theoretical physics interpretation of such problems, has been highly
successful in solving hard computational problem~\cite{MPZ}, far
beyond other existing approaches. In addition, the algorithm
facilitated theoretical studies of the corresponding physical system
and contributed to our understanding of it~\cite{MZ_PRE}.

Inspired by the extension of BP to SP we have extended the approach
of~\cite{Kabashima_CDMA}, designed for inference in densely connected
systems, in a similar manner to include an average over multiple pure
states. In this article we derive this extension, apply it to the
problem of CDMA signal detection~\cite{Kabashima_CDMA} and devise a
practical algorithm based on insight gained from the analysis. The
approach is general and can be applied to a broad range of
inference problems. However, for giving a specific example and highlighting
the advantages with respect to the original
method~\cite{Kabashima_CDMA} we will focus here on the application to
CDMA signal detection.

Multiple access communication refers to the transmission of multiple
messages to a single receiver. The scenario we study here is that of
\( K \) users transmitting independent messages over an additive white
Gaussian noise (AWGN) channel of zero mean and variance \( \sigma
_{0}^{2} \). Various methods are in place for separating the messages,
in particular Time, Frequency and Code Division Multiple
Access~\cite{CDMA_book}. The latter, is based on spreading the signal
by using \( K \) individual random binary spreading codes of spreading
factor \( N \).  We consider the large-system limit, in which the
number of users $K$ tends to infinity while the system load
$\beta\equiv K/N$ is kept to be ${\cal O} (1)$. We focus on a CDMA system
using binary phase shift keying (BPSK) symbols and will assume the
power is completely controlled to unit energy. The received
aggregated, modulated and corrupted signal is of the form:
\[ y_{\mu }=\frac{1}{\sqrt{N}}\sum ^{K}_{k=1}s_{\mu
k}b_{k}+\sigma _{0}n_{\mu }\]
where \( b_{k} \) is the bit transmitted by user \( k \), \( s_{\mu k}
\) is the spreading chip value, \( n_{\mu } \) is the Gaussian noise
variable drawn from \( {\mathcal{N}}\left( 0,1 \right) \), and \(
y_{\mu } \) the received message. The goal is to get an accurate
estimate of the vector \( \mathbf{b} \) for all users given the
received message vector \( \mathbf{y} \) by approximating the
posterior \( P(\mathbf{b}|\mathbf{y}) \). A method for obtaining a
good estimate of the posterior probability in the case where the noise
level is accurately known has been presented
in~\cite{Kabashima_CDMA}. However, the calculation is based on finding
a single solution and is therefore bound to fail, as have been
observed,when the solution space becomes fragmented, for instance when
the noise level is unknown, a case that arguably corresponds to
replica symmetry breaking.

The reason for the failure in this case can be qualitatively
understood by the same arguments as in the case of sparse graphs;
the existence of competing solutions results in inconsistent
messages and prevent the algorithm from converging to an accurate
estimate.  An improved solution can therefore be obtained by
averaging over the different solutions, inferred from the same
data, in a manner reminiscent to the SP approach, only that the
messages in the current case are more complex.

\begin{figure}
\begin{center}
\epsfig{file=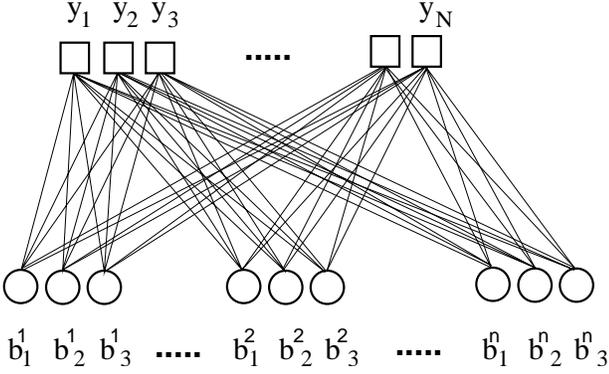,angle=0,width=85mm} \caption{Replicated
solutions \( {\mathbb{B}}\!=\!\left( \mathbf{b}_{1},\,
\mathbf{b}_{2},..,\mathbf{b}_{K}\right) \) given data.}
\vspace*{-0.5cm} \label{fig1}
\end{center}
\end{figure}

Figure~\ref{fig1} shows the detection problem we aim to solve as a
bipartite graph where \( {\mathbb{B}}=\left( \mathbf{b}_{1},\,
\mathbf{b}_{2},\, \ldots ,\, \mathbf{b}_{K}\right) \) the set of
bit vectors, \( \mathbf{b}_{k}=\left( b_{k}^{1},\, b_{k}^{2},\,
\ldots ,\, b_{k}^{n}\right) \), where \( n \) is the solution
(replica) index.

Using Bayes rule one obtains the BP equations:
\begin{eqnarray}
P^{t+1}\!\left(y_{\mu }|\mathbf{b}_{k},\left\{ y_{\nu \neq \mu }\right\} \right)
& \!=\! & \widehat{a}_{\mu k}^{t+1}\sum_{\mathbf{b}_{l\neq k}} \!
 P\left( y_{\mu }|{\mathbb{B}}\right) \nonumber\\&\,&\times\prod _{l\neq k}P^{t}
 \left( \mathbf{b}_{l}|\left\{y_{\nu\neq\mu }\right\} \right) \label{bp1} \nonumber \\
P^{t}\left( \mathbf{b}_{l}|\left\{ y_{\nu \neq \mu }\right\} \right)
&\!=\! & a_{\mu k}^{t} \prod _{\nu \neq \mu }P^{t}\left( y_{\nu }|\mathbf{b}_{l}\, ,\,
\left\{ y_{\sigma \neq \nu }\right\} \right) \label{bp2}
\end{eqnarray}
where \( \widehat{a}_{\mu k}^{t+1} \) and \( a_{\mu k}^{t} \) are
normalization constants.  For calculating the posterior
\begin{equation}
P\left( {\mathbb{B}}\, |\, \mathbf{y}\right) =\frac{\prod _{\mu
=1}^{N}P\left( y_{\mu }\, |\, {\mathbb{B}}\right) }{{\textsf
{Tr}}_{\left\{ {\mathbb{B}}\right\} }\prod _{\mu =1}^{N}P\left(
y_{\mu }\, |\, {\mathbb{B}}\right) },
\end{equation}
an expression representing the likelihood is required and is easily
derived from the noise model (assuming zero mean and variance
$\sigma^2$)
 \begin{equation}
\label{supp}
P \left( y_{\mu }\, |\, \mathbb{B} \right) =\frac{1}{\sqrt{2\pi
\sigma ^{2}}} \exp \left\{ -\frac{\left( \mathbf{y}_{\mu }-\Delta
_{\mu }\right)^{\tT} {\mathbb I}\left( \mathbf{y}_{\mu
}-\Delta _{\mu }\right) }
 {2 \sigma^{2}}      \right\} ,
\end{equation}
where \( \mathbf{y}_{\mu }=y_{\mu }\mathbf{u} \) and \(
\mathbf{u}^{\tT}\equiv \stackrel{{n}}{\overbrace{\left(
1,\, 1,\, \cdots ,\, 1\right) }} \)
\[
\Delta _{\mu }\equiv\frac{1}{\sqrt{N}}\sum ^{K}_{k=1}s_{\mu
k}\mathbf{b}_{k}.\]
An explicit expression for inter-dependence between solutions is
required for obtaining a closed set of update equations. We
assume a dependence of the form
 \begin{equation}
\label{message}
 P^{t}\left( \mathbf{b}_{k}\, |\, \left\{ y_{\nu \neq \mu
}\right\} \right) \propto \exp \left\{ \mathbf{h}_{\mu
k}^{t\tT}\,
\mathbf{b}_{k}+\frac{1}{2}\mathbf{b}^{\tT}_{k}{\mathbb
Q}^{t}_{\mu k}\, \mathbf{b}_{k}\right\} ,
\end{equation}
where \( \mathbf{h}_{\mu k}^{t} \) is a vector representing an external
field and \( {\mathbb Q}^{t}_{\mu k} \) the matrix of
cross-replica interaction. Furthermore, we assume the following
symmetry between replica:
\begin{eqnarray}\left( {\mathbb Q}^{t}_{\mu k}\right)
^{{\rm{ab}}} &=&\delta ^{\, {\rm{ab}}}\, q_{\mu
k}^{t}+\left( 1-\delta ^{\, {\rm{ab}}}\right) \, p_{\mu k}^{t}
\\ \mathbf{h}_{\mu k}^{t}&=&h_{\mu k}^{t}\mathbf{u} \nonumber .
\end{eqnarray}
An  expression for equation (\ref{message}) immediately follows
\begin{eqnarray}
P^{t}\!\left(\mathbf{b}_{k}|\left\{ y_{\nu \neq \mu }\right\}
\right) &=&[\mathcal{Z}^{t}_{\mu k}]^{-1}\\
&\,&\times\exp \left\{ h_{\mu k}^{t}\sum
_{{\rm{a}}=1}^{n}b^{{\rm{a}}}_{k}\!+\!\frac{1}{2}p_{\mu
k}^{t}\left( \sum _{{\rm{a}}=1}^{n}b^{{\rm{a}}}_{k}\right)
^{2}\!\right\}\nonumber ,\end{eqnarray}
where \(\mathcal{Z}^{t}_{\mu k}\) is a normalization
 constant.

We expect the free energy obtained from the well behaved
distribution \( P^{t} \) to be self-averaging, thus\[ \lim
_{n\rightarrow \infty }\frac{1}{n}\log \left(
\overline{\mathcal{Z}^{t}_{\mu k} }\right) =\lim _{n\rightarrow
\infty }\frac{1}{n}\log \left( \mathcal{Z}^{t}_{\mu k}\left(
h_{\mu k}^{t},\, q_{0},\, p_{0}\right) \right) ,\]
where the
sub-index 0 represents the mean value of the parameters when
extracted for some suitable distributions and the overline
represents the mean value of the partition function over such
distributions.

To obtain the scaling behavior of the various parameters we
calculate $\mathcal{Z}\left( h,\, q,\, p\right)$ explicitly,
assuming the parameters $q$ and $p$ are taken from normal
distributions \( \mathcal{N}_{q}\left( q_{0},\sigma
_{q}^{2}\right)  \) and \( \mathcal{N}_{p}\left( p_{0},\sigma
_{p}^{2}\right)  \). After a long calculation one obtains the
following scaling: \( h\!\sim\! \mathcal{O}\left( 1\right)  \), \(
q_{0}\!\sim \!\mathcal{O}\left( 1\right)  \), \( p_{0}\!\sim\!
\mathcal{O}\left( n^{-1}\right)  \), \( \sigma _{q}^{2}\!\sim\!
\mathcal{O}\left( n^{-1}\right)  \), and \( \sigma
_{p}^{2}\!\sim\! \mathcal{O}\left( n^{-3}\right)  \). In the
remainder of the paper
 we will rescale the off-diagonal elements of \( {\mathbb
Q}_{\mu k}^{t} \)  to \( g_{\mu k}^{t}/n \), where \( g_{\mu
k}^{t}\!\sim\! \mathcal{O}\left( 1\right)  \).

The  marginalized posterior at time \emph{t} takes the form
\begin{eqnarray} \label{pp} P^{t}\!\left( \mathbf{b}_{k}|\left\{ y_{\nu \neq \mu }\right\} \right) &\!=\!& \!\frac{\int _{-\infty
}^{\infty }dx\, \exp \left\{\!-n\frac{\left( x-h_{\mu k}^{t}\right)
^{2}}{2g_{\mu k}^{t}}\!+\!x\sum
_{{\rm{a}}=1}^{n}\!b_{k}^{{\rm{a}}}\right\}}{\int _{-\infty
}^{\infty }dx\, \exp \left\{ -n\Phi \left( x;\, h_{\mu k}^{t},\,
g_{\mu k}^{t}\right) \right\} } \nonumber \\
\Phi \left( x;\, h_{\mu k}^{t},\, g_{\mu k}^{t}\right) &\!=\!&
-\frac{\left( x-h_{\mu k}^{t}\right) ^{2}}{2g_{\mu k}^{t}}\!+\!\ln
\left( \cosh (x)\right) \ .
\end{eqnarray}
To find the dominant solutions in the case of large $n$ one studies
the maxima of \( \Phi \left( x;\, h,\, g\right) \). One identifies
regimes with a single and double peaks, depending on the values of $h$
and $g$ (full details will be given elsewhere); the main contribution
comes from a regime where \( g^{t}_{\mu k}>1 \) and \( 0<h^{t}_{\mu
k}/g^{t}_{\mu k}\ll 1 \), where \( \Phi \left( x;\, h,\, g\right) \)
takes the form of an almost symmetric pair of Gaussians located at
\begin{equation}
x^{t}_{\pm,\mu k}  \simeq  \pm x^{t}_{0,\mu k}+\frac{g^{t}_{\mu k}}
{g^{t}_{\mu k}+\left( x^{t}_{0,\mu k}\right) ^{2}- \left(
g^{t}_{\mu k}\right)^{2}}h^{t}_{\mu k} \ ,
\end{equation}
where \( \pm x_{0} \) are the positions of the peaks at zero
field.

To calculate correlation between replica we expand $P\left( y_{\mu
}\, |\, {\mathbb B}\right)$ in the large \emph{N} limit
(Eq.~\ref{supp}), as in~\cite{Kabashima_CDMA}, to obtain
\begin{eqnarray} \label{gaprox} P\left(
y_{\mu }\, |\, {\mathbb B}\right) &\simeq& \left( \frac{\exp \left(
-1/2N\sigma ^{2}\right) }{\sqrt{2\pi \sigma ^{2}}}\right) ^{n}\\
&\,&\times\exp
\left\{ -\frac{\left( \mathbf{y}_{\mu }-\Delta _{\mu k}\right)
^{{\sf T}}{\mathbb I}\left( \mathbf{y}_{\mu }-\Delta _{\mu
k}\right) }{2\sigma ^{2}}\right\} \nonumber\\
&\,&\times\left[ 1+\frac{s_{\mu
k}}{\sqrt{N}\sigma ^{2}}\left( \mathbf{y}_{\mu }-\Delta _{\mu
k}\right) ^{{\sf T}}\mathbf{b}_{k} \ , \right]\nonumber
\end{eqnarray}
where $ \Delta _{\mu k}=\frac{1}{\sqrt{N}}\sum _{l\neq k}s_{\mu
l}\mathbf{b}_{l}$.

For large \emph{n} and small field we obtain the following
\begin{eqnarray}
\left\langle b_{k}^{{\rm a}}\right\rangle &\!=\!& {\textsf
{Tr}}_{\left\{ \mathbf{b}_{k}\right\} }
P^{t}\left(\mathbf{b}_{k}|\left\{ y_{\nu \neq \mu}\right\}\right)
b_{k}^{{\rm{a}}}\nonumber\\
 &\simeq&  a_{+,\mu k}^{t}\tanh\left(x^{t}_{+,\mu
k}\right)\!+\!a_{-,\mu k}^{t}\tanh \left( x^{t}_{-,\mu k}\right)
\label{meanbnew} \nonumber \\ 
\left\langle b_{k}^{{\rm a}}b_{k}^{{\rm
b}}\right\rangle &\!=\!& {\textsf
{Tr}}_{\left\{\mathbf{b}_{k}\right\}}P^{t}\left(
\mathbf{b}_{k}|\left\{ y_{\nu \neq \mu}\right\}\right)
b_{k}^{{\rm{a}}}b_{k}^{{\rm{b}}}\nonumber\\
&\simeq& a_{+,\mu
k}^{t}\tanh\left(x^{t}_{+,\mu k}\right)^{2}+a_{-,\mu k}^{t}\tanh
\left( x^{t}_{-,\mu k}\right)^{2},\label{meanbbnew} \nonumber \\
\left\langle b_{k}^{{\rm{a}}}b_{l}^{{\rm{b}}}\right\rangle
&\!=\!&\left\langle b_{k}^{{\rm{a}}}\right\rangle \left\langle
b_{l}^{{\rm{b}}}\right\rangle .
\end{eqnarray}
where \( m_{\mu k}^{t}\equiv \tanh\left(x^{t}_{0,\mu k}\right)
\equiv x^{t}_{0,\mu k}/g^{t}_{\mu k}\), and
\begin{equation}
a_{\pm,\mu k}^{t} \simeq  \frac{\exp \left\{ \mp nm_{\mu
k}^{t}h_{\mu k}^{t}\right\} } {\exp \left\{ nm_{\mu k}^{t}h_{\mu
k}^{t}\right\} +\exp \left\{ -nm_{\mu k}^{t}h_{\mu k}^{t}\right\}
} \  . \label{a+-}
\end{equation}
Using Eqs. (\ref{meanbnew}) we calculate the first two cummulants
of the elements of \( \Delta _{\mu k} \):
\begin{eqnarray}
\left\langle \Delta _{\mu k}^{{\rm a}}\right\rangle
&=&\frac{1}{\sqrt{N}}\sum _{l\neq k}s_{\mu l}m^{t}_{\mu l} \\
\left( \chi _{\mu k}^{t}\right) ^{{\rm ab}} & \equiv &
\left\langle \Delta _{\mu k}^{{\rm a}}\Delta _{\mu k}^{{\rm b}}
\right\rangle -\left\langle \Delta _{\mu k}^{{\rm a}}\right\rangle
\left\langle \Delta _{\mu k}^{{\rm b}}\right\rangle \nonumber \\
 & = & \delta ^{{\rm ab}}\beta \, \left( 1-Q_{\mu k}^{t}\right)
 +\left( 1-\delta ^{{\rm ab}}\right) \beta \, R_{\mu k}^{t},
 \nonumber
 \label{correl}
\end{eqnarray}
where $Q_{\mu k}^{t}$ and $R_{\mu k}^{t}$ can be approximated
using the law of large numbers as
\begin{eqnarray*}
Q_{\mu k}^{t} & \equiv & \frac{1}{K}\sum
_{l\neq k} \left( a_{+,\mu k}^{t}\tanh \left( x^{t}_{+,\mu
k}\right) +a_{-,\mu k}^{t}\tanh
\left( x^{t}_{-,\mu k}\right) \right) ^{2}\\
 & \simeq  & \frac{1}{K}\sum _{l\neq k}\left( m^{t}_{\mu k}\right) ^{2}\label{Q} \\
R_{\mu k}^{t} & \equiv & \frac{1}{K}\sum _{l\neq k}a_{+,\mu k}^{t}a_{-,\mu k}^{t}
\left( \tanh \left( x^{t}_{+,\mu k}\right) -\tanh \left( x^{t}_{-,\mu k}\right)
\right) ^{2}\\
 & \simeq  & \frac{4}{K}\sum _{l\neq k}a_{+,\mu k}^{t}a_{-,\mu k}^{t}
 \left( m^{t}_{\mu k}\right) ^{2} \equiv \frac{1}{n}\Upsilon _{\mu k}^{t}.\label{R}
\end{eqnarray*}
The definition of $\Upsilon _{\mu k}^{t}$ relies on the expected
scaling of the off-diagonal terms of the matrix \( \chi _{\mu
k}^{t} \).

Thus, we expect the variables \( \Delta _{\mu k} \) to obey a
Gaussian distribution defined in Eqs.(\ref{correl}). The mean
value of \( b_{k}^{{\rm a}} \) at time \( t\!+\!1 \) is then given
by:
\begin{eqnarray}
\label{mhat} \widehat{m}_{\mu k}^{t+1}\!&=&\!\left(\sigma ^{2}\!+\!\beta
\left(1\!-\!Q_{\mu k}^{t}\right)\!+\!\beta\Upsilon_{\mu
k}^{t}\right)^{-1}\nonumber\\&\,&\left(\frac{y_{\mu }\mathbf{s}_{\mu
}}{\sqrt{N}}-\beta\left({\mathbb P}_{\mu }\!-\!{\mathbb
I}/K\right)\mathbf{m}^{t}_{\mu }\right)_{k},
\end{eqnarray}
where ${\mathbb P}_{\mu } \equiv (1/K) s_{\mu k}
s_{\mu l})$ and ${\mathbb I} \equiv \delta_{kl}
$, respectively. We assume that the macroscopic variables
are self averaging and omit the $\mu, k$ indices.

The main difference between Eq. (\ref{mhat}) and the equivalent
equation in~\cite{Kabashima_CDMA} is the emergence of an extra term in
the prefactor, $\beta \Upsilon^{t}$, reflecting correlations between
different solutions groups (replica). To determine this term we
optimize the choice of $\Upsilon^{t}$ by minimizing the bit error at
each time step.  Following~\cite{Kabashima_CDMA} we define
\begin{eqnarray}
M^{t} &\!\equiv\!& \frac{1}{NK}\sum ^{N}_{\mu =1} \sum
^{K}_{k=1}b_{k}m_{\mu k}^{t} \label{mt1}
  \!= \! \int \mathcal{D}z\, \tanh \left(\sqrt{F^{t}}z\!+\!E^{t}\right) \label{mt2} \\
Q^{t} &\!\equiv\!& \frac{1}{NK}\sum ^{N}_{\mu =1}\sum
^{K}_{k=1} \left(b_{k}m_{\mu k}^{t}\right)^{2}\label{qt1}
  \!=\!\int \mathcal{D}z\, \tanh ^{2}\left(\sqrt{F^{t}}z\!+\!E^{t}\right) ,\label{qt2}\nonumber
\end{eqnarray}
where \( \mathcal{D}z\equiv {\rm d}z\, \exp \left[ -z^{2}/2\right]
/\sqrt{2\pi } \) and
\begin{eqnarray}
E^{t+1} & \!\equiv\! & \frac{1}{K}\sum ^{N}_{\mu =1}\,
\sum ^{K}_{k=1}b_{k}\widehat{m}_{\mu k}^{t+1}  \!= \! \frac{1}{\sigma ^{2}+\beta \left( 1-Q^{t}+\Upsilon ^{t}\right) } \nonumber \\
F^{t+1} & \!\equiv\! & \sum ^{N}_{\mu =1}\left[ \frac{1}{K}\, \sum
^{K}_{k=1}\left( b_{k}\widehat{m}_{\mu k}^{t+1}\right)
^{2}-\frac{1}{K^{2}}
\left( \sum ^{K}_{k=1}b_{k}\widehat{m}_{\mu k}^{t+1}\right) ^{2}\right] \nonumber\\
 & \!\simeq\!  & \frac{1}{K}\sum ^{N}_{\mu =1}\widehat{\mathbf{m}}^{t+1}_{\mu }\cdot
  \widehat{\mathbf{m}}^{t+1}_{\mu }\label{ft2} \\
 & \!=\! & \left[ \beta \left( 1-2M^{t}+Q^{t}\right) +\sigma _{0}^{2}\right]
 \left( E^{t+1}\right) ^{2} \nonumber \ ,
\end{eqnarray}
To obtain the bit error rate:
\begin{equation}
P_{b}^{t} \equiv  \frac{1}{2K}\sum ^{K}_{k=1}\left( b_{k}-{\rm sgn}\left( m_{k}^{t}
\right) \right) \\
 =  \int _{-\infty }^{-E^{t}/\sqrt{F^{t}}}\mathcal{D}z\label{ptb}
\end{equation}
\begin{equation}
\mbox{with~~} m_{k}^{t}\simeq \tanh \left( \sum ^{N}_{\mu =1}\widehat{m}_{\mu
k}^{t}\right) \label{mupdate}.
\end{equation}

Optimizing  $P_{b}^{t}$ with respect to $\Upsilon^{t}$
one obtains straightforwardly that $ E^{t}=F^{t}$ and \(
Q^{t}=M^{t} \). In principle, the optimization can be done 
globally~\cite{SR_PRL} but is of a limited practical value.

This implies that $ \Upsilon ^{t} \!=\! (\sigma _{0}^{2}\!-\!\sigma
^{2})/\beta$  is just a constant. However, it holds the key
to obtaining accurate inference results. If the noise estimate is
identical to the true noise the term vanishes and one retrieves
the expression of~\cite{Kabashima_CDMA}; otherwise, an estimate of
the difference between the two noise values is required for
computing \( E^{t} \).

From equation (\ref{ft2}) one obtains:
\begin{eqnarray}
\label{expfinal}
E^{t+1}&\!\!\simeq\!& \frac{1}{K}\sum ^{N}_{\mu
=1}\widehat{\mathbf{m}}^{t+1}_{\mu }
\cdot \widehat{\mathbf{m}}^{t+1}_{\mu } \\
 &\!=\!& \left[\frac{1}{\sigma ^{2}\!+\!\beta\left(1\!-\!Q^{t}\!+\!\Upsilon^{t}\right)}\right]^{2}
 \left[\frac{1}{N}\!\sum^{N}_{\mu=1}y_{\mu }^{2}\!-\!\beta \left(2M^{t}\!-\!Q^{t}\right) \right] \nonumber \\
 &\!=\!& \left( E^{t+1}\right)^{2}\left[\frac{1}{N}\sum^{N}_{\mu=1}y_{\mu}^{2}\!-\!\beta Q^{t}\right]
  \!=\! \left[\frac{1}{N}\sum^{N}_{\mu=1}y_{\mu }^{2}\!-\!\beta Q^{t}\right]^{\!-1\!}  \nonumber .
\end{eqnarray}
This enables us to rewrite Eq.(\ref{mhat}) as
\begin{eqnarray}
\label{mhatfin}
\widehat{m}_{\mu k}^{t+1} & = & A^{t}\, \left( \frac{y_{\mu }\mathbf{s}_{\mu }}{\sqrt{N}}-\beta \, \left( {\mathbb P}_{\mu }-K^{-1}{\mathbb I}\right) \, \mathbf{m}^{t}_{\mu }\right) _{k} \\
A^{t} & = & \left\{ \frac{1}{N}\sum ^{N}_{\mu =1}y_{\mu
}^{2}-\beta Q^{t}\right\} ^{-1} \nonumber 
\end{eqnarray}
{\em  where no estimate on \( \sigma _{0} \) is required}.

\begin{figure}
\begin{center}
\epsfig{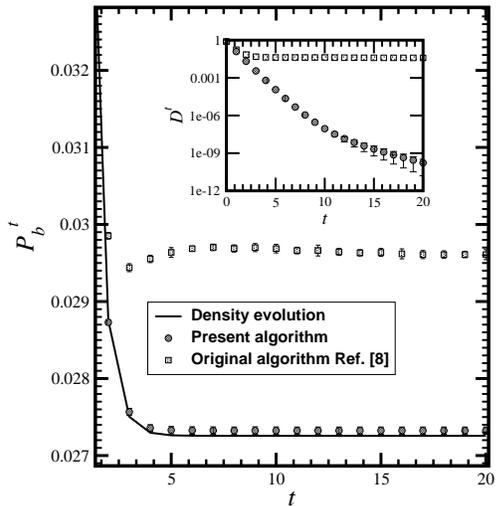} \end{center}\caption{Error probability
of the inferred solution evolving in time.  The system load
$\beta=0.25$, true noise level \( \sigma _{0}^2 = 0.25 \) and
estimated noise \( \sigma^2 = 0.01 \). Squares represent results of
the original algorithm~\cite{Kabashima_CDMA}, solid line the dynamics
obtained from our equations; circles represent results obtained from
the suggested {\em practical} algorithm. Variances are smaller than
the symbol size.  In the inset, $D^t$, a measure of convergence in the
obtained solutions, as a function of time; symbols are as in the main
figure.}  \label{fig2}

\end{figure}
The inference algorithm requires an iterative update of
Eqs.(\ref{expfinal},\ref{mhatfin},\ref{mupdate}) and
converges to a reliable estimate of the signal, with no need for
an accurate prior information of the noise level. The
computational complexity of the algorithm is of ${\cal O}(K^2)$.

To test the performance of our algorithm we carried out a set of
experiments of CDMA signal detection problem under typical
conditions. Error probability of the inferred signals has been
calculated for a system load of $\beta\!=\!0.25$, where the true noise
level is \( \sigma _{0}^2 \!=\! 0.25 \) and the estimated noise is \(
\sigma^2 \!=\! 0.01 \), as shown in Figure~\ref{fig2}.  The solid line
represents the expected theoretical results (density evolution),
knowing the exact values of \( \sigma _{0}^2\) and \( \sigma^2 \),
while circles represent simulation results obtained via the suggested
{\em practical} algorithm, where no such knowledge is assumed. The
results presented are based on $10^5$ trials per point and a system
size $N\!=\!2000$ and are superior to those obtained using the
original algorithm~\cite{Kabashima_CDMA}.

 Another performance measure one should consider is  \[
D^{t}\equiv\frac{1}{K}\left(
\mathbf{m}^{t}-\mathbf{m}^{t-1}\right) \cdot \left(
\mathbf{m}^{t}-\mathbf{m}^{t-1}\right) , \] that provides an
indication to the stability of the solutions obtained. In the
inset of Figure~\ref{fig2} we see that results obtained from our
algorithm show convergence to a reliable solution in stark
contrast to the original
 algorithm~\cite{Kabashima_CDMA}. The physical interpretation of
 the difference between the two results is assumed to be related
 to a replica symmetry breaking phenomena.

In summary, we present a new algorithm for using belief
propagation in densely connected systems that enables one to
obtain reliable solutions even when the solution space is
fragmented. It represents an extension to existing algorithms of
that type which is reminiscent to the extension of BP to SP. The
algorithm has been tested on the signal detection problem in CDMA
and has provided superior results to other existing
algorithms~\cite{Kabashima_CDMA,Kabashima_new}. Further research
is required to fully determine the potential of the new algorithm.


Support from the EU FP-6 EVERGROW IP is gratefully acknowledged.



\begin{thebibliography}{99}
\bibitem{Kabashima_new} In preparing the manuscript we have found that
a similar activity is being carried out by Yoshiyuki Kabashima at
TITECH. However, in the absence of details we cannot comment on his
approach.
\end{thebibliography}
\end{document}